\documentclass{elsarticle}
\usepackage[margin=3.4cm]{geometry}
\usepackage{latexsym}
\usepackage{longtable}
\usepackage{float}
\usepackage[latin2]{inputenc}
\usepackage{graphicx}
\usepackage{ae}
\usepackage{aecompl}
\usepackage{setspace}
\usepackage{amsmath}
\usepackage{amssymb}
\usepackage{amsthm}
\usepackage[hidelinks]{hyperref}

\makeatletter
\def\ps@pprintTitle{%
	\let\@oddhead\@empty
	\let\@evenhead\@empty
	\def\@oddfoot{\centerline{\thepage}}%
	\let\@evenfoot\@oddfoot}
\makeatother

\begin{document}

\title{New Graphs at the braingraph.org Website for Studying the Aging Brain Circuitry}

\author[p]{Bálint Varga}
\ead{balorkany@pitgroup.org}
\author[p,u]{Vince Grolmusz\corref{cor1}}
\ead{grolmusz@pitgroup.org}
\cortext[cor1]{Corresponding author}
\address[p]{PIT Bioinformatics Group, Eötvös University, H-1117 Budapest, Hungary}
\address[u]{Uratim Ltd., H-1118 Budapest, Hungary}

\date{}

\begin{abstract}
Human braingraphs or connectomes are widely studied in the last decade to understand the structural and functional properties of our brain. In the last several years our research group has computed and deposited thousands of human braingraphs to the braingraph.org site, by applying public structural (diffusion) MRI data from young and healthy subjects. Here we describe a recent addition to the {\tt braingraph.org} site, which contains connectomes from healthy and demented subjects between 42 and 95 years of age, based on the public release of the OASIS-3 dataset. The diffusion MRI data was processed with the Connectome Mapper Toolkit v.3.1.
We believe that the new addition to the braingraph.org site will become a useful resource for enlightening the aging circuitry of the human brain in healthy and diseased subjects, including those with Alzheimer's disease in several stages.
\end{abstract}

\maketitle

\section*{Introduction}

Connectomes or braingraphs are widely used today in the study of human brain. Two main approaches are the functional MRI-based functional connectomes and the diffusion MRI-based structural or anatomical connectomes. Here we restrict our attention to the latter. 

The diffusion MR imaging is capable of describing the macroscopic diffusion directions of water molecules in the neurons of the brain, and since in the white matter most water molecules move in the direction of the axons of neurons, it is possible to uncover the orbits of the axon bundles, which connect gray matter areas in the cortex and the subcortex.  If the gray matter areas of the cortex and subcortex are labeled by their anatomical names, then we can build braingraphs or connectomes as follows: their vertices correspond to the anatomically labeled gray matter areas (frequently called Regions Of Interests, ROIs), and two such nodes are connected by an edge if a diffusion magnetic resonance imaging based tractography workflow \cite{Besson2014a} finds an axon-bundle between the two areas.  

This way the imaging information of the MRI is translated into a discrete graph structure, and therefore, we can make use of the rich and well-developed resources of graph theory, introduced in 1741 by the famous mathematician Leonhard Euler \cite{Eulera}, and grown into maturity in the 20th and 21st centuries by hundreds of great mathematicians.  

Our research group has computed and published several undirected and directed braingraph sets with upto 1015 vertices for each brain \cite{Kerepesi2016b,Szalkai2015a,Szalkai2016,Kerepesi2015b, Szalkai2016d} from different Data Releases of the Human Connectome Project \cite{McNab2013}. The computed graphs were made available at the site \url{https://braingraph.org}, and were applied in several structural studies of the young and healthy human brain \cite{Szalkai2015,Kerepesi2015a, Szalkai2016d, Kerepesi2016, Szalkai2016c, Szalkai2017c, Szalkai2016e, Szalkai2016a,Fellner2017,Fellner2019,Fellner2018}. 

Since the Human Connectome Project published the MRI data of young and healthy subjects only, the above listed resources will not help in studying the aging of the human brain. Since the expected age of humans are increasing almost everywhere in the world, the age-related neurodegenerative diseases, especially dementias cause a big burden on the health care systems worldwide.

Dementia is rare before age 60, but doubled by every five years of age thereafter \cite{Bermejo-Pareja2008, Carlo2002}, it affects about 40 percent of those over 90, and up to 20 percent of those between 75 and 84 \cite{Wortmann2012, war2009}. By the most recent data of WHO, 55 million people has dementia worldwide, and 10 million new cases are diagnosed in each year. The cost of dementia is estimated to be 1.3 x $10^{15}$ \$ US. The most common cause of dementia is Alzheimer's disease (AD). The earliest symptoms of AD include forgetfulness; disorientation to time or place; and difficulty with concentration, calculation, language, and judgment. As the disease progresses, some patients have severe behavioural disturbances and may even become psychotic. In the final stages, the affected individual is incapable of self-care and becomes bed-bound. 

Aging-related anatomical changes are well-known for a long time: Numerous organs lose weight and volume, like the muscles, the lung, the kidneys, the liver. Overall bone mass decreases. The skin becomes thinner and stiffer, the cartilage of the skeletal joints degenerate leading to painful movement. Thickening and increased rigidity of the walls of arteries and deposits in them cause the cardiovascular system several malfunctions.  The knowledge of those age-related changes has led to hundreds of treatments, novel drugs, life improving surgical procedures which make possible the healthier aging and longer self-sufficiency in the life.

We know much less of the anatomical changes in the connections in the human brain during aging. Numerous studies described brain volumetric decrease in cortical areas in aging, e.g., \cite{Bigler1997,Asken2023}. 

Here we present a public resource, containing the data of 696 subjects of ages between 42 and 95 years.  Some subjects were MRI scanned more than once in a several years span; therefore,  we present 975 braingraphs in total. Each of these 975 braingraphs is computed in 5 resolutions. This dataset may be an important source for studying the aging of human brain connections in healthy and demented subjects. Describing those changes adequately may also lead to revolutionary therapies and novel drugs which improve the function of the aging brain. 

In what follows, we describe the method of the braingraph construction, and also the resulting dataset available under Section A at \url{https://braingraph.org/cms/download-pit-group-connectomes/}.

\section*{Methods}

\subsection*{Data Source}

The data source is the public release of OASIS-3 dataset \cite{LaMontagne2019}. The resource contains MRI and PET imaging recordings together with rich clinical data, from 1098 subjects between age 42 and 95 years, over a 15 year time span.  Diffusion MRI data of 1472 sessions were recorded using Siemens 3T scanners of two models: TIM Trio
3T, and BioGraph mMR PET-MR 3T at the Washington University Knight Alzheimer Disease Research Center \cite{LaMontagne2019}.

\subsection*{Computational Workflow}

The computation of braingraphs involves the identification of anatomically labeled gray matter areas (parcellation) and the computation of axonal fiber tracts (or streamlines) which connect those gray matter areas (tractography). The resulting braingraph has a vertex-set, corresponded to the anatomically labeled gray matter areas, and two such nodes are connected by an edge if the tractography step finds at least one axonal fiber, which connects them. 

For the computation we have applied the Connectome Mapper Tool Kit v.3.1., abbreviated CMP3.1 \cite{Tourbier2022,TourbierZenodo}, with probabilistic tractography. From each MRI data set we have prepared 5 graphs with different resolutions with 124, 170, 272, 502 and 1058 vertices, according to Lausanne2018 brain parcellations.

\section*{Discussion and Results}

 The OASIS-3 dataset \cite{LaMontagne2019} contains the multimodal MRI and PET data of more than 1000 subjects from 42 through 95 years of age, with good annotations of the psychiatric and general health status. We were able to process the data of 696 subjects from OASIS-3; the remaining subjects have one or more missing or erroneous files, which prevented their processing with the CMP3.1 workflow. More than one MRI were processed from numerous subjects, therefore, we have computed 975 graphs from the diffusion MRI data: one dataset from 482 subjects, 2 datasets from 156 subjects, 3 datasets from 52 subjects, 4 datasets from five subjects, 5 datasets from 1 subject.

\subsection*{Description of edge and node attributes}

In the braingraphs deposited at the \url{https://braingraph.org} site, the nodes and the edges have several attributes, detailed below:

\subsection*{Node attributes}

The node attributes in the GraphML files include the following values:
\begin{itemize}
\item dn\_region can be cortical or subcortical

\item dn\_position\_x, dn\_position\_y, dn\_position\_z the coordinates of a vertex. In a few cases, mostly in the substructures of the hippocampus, their values are missing if the substructure is not identified reliably. In that case the NAN abbreviation (not a number) is given there.

\item dn\_name the corresponding anatomical name; it is either identical to or a refinement of the fsname attribute.

\item dn\_hemisphere  left or right 

\item dn\_fsname  the corresponding anatomical area name of the node in FreeSurfer.

\item dn\_multiscaleID  numerical node ID 

\end{itemize}

\subsection*{Edge attributes}

The edge attributes in the GraphML files include the following quantities:

\begin{itemize}
\item number\_of\_fibers, corresponding to an edge. 

\item fiber\_length mean, median and standard deviation (std) of the fiber lengths in mm, corresponding to an edge. 

\item fiber\_density, normalized\_fiber\_density, fiber\_proportion:  Since fibers (or streamlines) can not always be tracked reliably in tractography algorithms, and since fibers may start or end erroneously in white matter during tractography (which is possible only in gray matter anatomically), some authors prefer to use fiber density quantities instead or besides of fiber numbers \cite{Zhang2022}.
\item 

The following quantities are related to the image reconstruction method SHORE: Simple harmonic oscillator based reconstruction and estimation \cite{Ozarslan2008, Koay2012,Ozarslan2013}:

\item shore\_rtop\_signal: RTOP: Return-to-Origin Probability \cite{Descoteaux2011}.

\item shore\_msd: MSD: Mean Squared Displacement, with standard deviation (std), mean and median values; \cite{Wu2007, Wu2008b}

\item shore\_gfa: GFA:  derived Generalized Fractional Anisotropy (GFA) with standard deviation (std), mean and median values \cite{Tuch2004};

\end{itemize}

\section*{Acknowledgments}
Data were provided in part by OASIS-3 Longitudinal Multimodal Neuroimaging: Principal Investigators: T. Benzinger, D. Marcus, J. Morris; NIH P50 AG00561, P30 NS09857781, P01 AG026276, P01 AG003991, R01 AG043434, UL1 TR000448, R01 EB009352. BV and VG were partially supported by the ELTE TKP 2021-NKTA-62 project.
\bigskip 

\section*{Data availability} The braingraphs are available under Section A at the site \url{https://braingraph.org/cms/download-pit-group-connectomes/}.

\noindent Conflict of Interest: The authors declare no conflicts of interest.

\section*{Author Contribution} BV constructed the image processing system, computed the braingraphs, and prepared the figure, VG has secured funding, initiated the study, analyzed data and wrote the paper.



\end{document}